\documentclass[]{aa}
\usepackage{graphicx}
\usepackage{txfonts}

\usepackage{natbib}
\bibpunct{(}{)}{;}{a}{}{,}
\begin{document}
   \title{The HeII Lyman alpha forest and the thermal state of the IGM}

   \subtitle{}

   \author{C. Fechner
          \inst{1}
          \and
          D. Reimers\inst{1}
          }

   \offprints{C. Fechner}

   \institute{Hamburger Sternwarte, Universit\"at Hamburg,
              Gojenbergsweg 112, 21029 Hamburg, Germany\\
              \email{[cfechner,dreimers]@hs.uni-hamburg.de}
             }

   \date{Received 16 October 2006; accepted 15 November 2006}

   \abstract{
Recent analyses of the intergalactic UV background by means of the \ion{He}{ii} Ly$\alpha$ forest assume that \ion{He}{ii} and \ion{H}{i} absorption features have the same line widths.
We omit this assumption to investigate possible effects of thermal line broadening on the inferred \ion{He}{ii}/\ion{H}{i} ratio $\eta$ and to explore the potential of intergalactic \ion{He}{ii} observations to constrain the thermal state of the intergalactic medium.
Deriving a simple relation between the column density and the temperature of an absorber based on the temperature-density relation $T = T_0 (1+\delta)^{\gamma -1}$ we develop a procedure to fit $T_0$, $\gamma$, and $\eta$ simultaneously by modeling the observed spectra with Doppler profiles.
In an alternative approach the temperature $T$ of an absorber, the \ion{He}{ii}/\ion{H}{i} ratio $\eta$, and the redshift scale of $\eta$ variations are estimated simultaneously by optimizing the Doppler parameters of \ion{He}{ii}.
Testing our procedure with artificial data shows that well-constrained results can be obtained only if the signal-to-noise ratio in the \ion{He}{ii} forest is $S/N \gtrsim 20$.
Additionally, ambiguities in the line profile decomposition may result in significant systematic errors.
Thus, it is impossible to give an estimate of the temperature-density relation  with the \ion{He}{ii} data available at present ($S/N \sim 5$).
However, we find that only 45\,\% of the lines in our sample favor turbulent line widths while the remaining lines are probably affected by thermal broadening.
Furthermore, the inferred $\eta$ values are on average about $0.05\,\mathrm{dex}$ larger if a thermal component is taken into account, and their distribution is 46\,\% narrower in comparison to a purely turbulent fit.
Therefore, variations of $\eta$ on a 10\,\% level may be related to the presence of thermal line broadening.
The apparent correlation between the strength of the \ion{H}{i} absorption and the $\eta$ value, which has been found in former studies assuming turbulent line broadening, essentially disappears if thermal broadening is taken into account.
In the redshift range $2.58 < z < 2.74$ towards the quasars
\object{HE~2347-4342} and \object{HS~1700+6416} we obtain $\eta \approx 100$ and slightly larger.
In the same redshift range the far-UV spectrum of HS~1700+6416 is best
and we estimate a mean value of $\log\eta = 2.11 \pm 0.32$ taking into account combined thermal and turbulent broadening.

   \keywords{cosmology: observations -- quasars: absorption lines -- 
     quasars: individual: HE~2347-4342, HS~1700+6416
               }
   }

   \titlerunning{The HeII Ly$\alpha$ forest and the thermal state of the IGM}
   \maketitle
%

\section{Introduction}

The temperature of the intergalactic medium (IGM) preserves a memory of the reionization history of the universe \citep[e.g.][]{theunsetal2002} and provides constraints on the nature of the reionizing sources \citep[e.g.][]{tittleymeiksin2006}.
The usual approach to measure the temperature considers the Doppler parameter distribution of Ly$\alpha$ forest absorbers \citep[e.g.][but see \citet{theunsetal2002c} for a different approach]{schayeetal1999, schayeetal2000, ricottietal2000, bryanmachacek2000, mcdonaldetal2001}.
Generally, there is a power law dependence between the column density $N_{\ion{H}{i}}$ and the lowest Doppler parameter at this column density, i.e.\ the cut-off $b_{\mathrm{min}}(N_{\ion{H}{i}})$.
Interpreting $b_{\mathrm{min}}$ as purely thermal, the temperature of the IGM can be estimated.
In practice, hydrodynamic simulations are required to transform the $b_{\mathrm{min}}$-$N_{\ion{H}{i}}$ dependency into a temperature-density relation \citep[e.g.][]{schayeetal1999}.
However, even for the narrowest lines the thermal component is not necessarily the only contribution to the line width.
Thus, the cut-off provides just an upper limit of the temperature.
Further broadening is introduced by the velocity field including peculiar differential motions and the Hubble flow.
The observed distribution of $b$-parameters is also related to the amplitude of the primordial density fluctuations as shown by \citet{huirutledge1999} using linear perturbation theory.

The approach of estimating the cut-off of the $b(N_{\ion{H}{i}})$ distribution to measure the temperature of the IGM is based on the concept of a well-defined power law relation between temperature and density.
\citet{huignedin1997} showed this relation to be valid in the low-density Ly$\alpha$ forest for uniform reionization models as a consequence of the interplay of photoionization heating and adiabatic cooling due to the cosmic expansion.
\citet{valageasetal2002} argued that the power law relation is still valid even if the ionizing UV background is inhomogeneous as long as the gas is in local ionization equilibrium.
The relation breaks at high densities since in the non-linear regime shock-heating due to gravitational processes becomes dominant introducing a second IGM component, the warm-hot intergalactic medium (WHIM) containing the bulk of the baryons at the present epoch \citep[e.g.][]{cenostriker1999, daveetal2001, valageasetal2002, kangetal2005}.
On the other hand, high-density gas may cool down in less than a Hubble time which is a precondition to enable star formation.
Thus, cool dense gas exists at $z \sim 2 - 3$ and even at lower redshifts \citep[e.g.][]{daveetal1999, valageasetal2002}.
As a consequence the temperature-density relation breaks at high densities ($\delta \sim 10$ or $N_{\ion{H}{i}} \sim 10^{15}\mathrm{cm}^{-2}$, respectively).
Above this threshold the gas temperature decreases with density \citep[for a summary of the phases of the IGM see e.g.\ Fig.\ 1 of][]{valageasetal2002}.

In agreement with the theoretical scenario \citet{misawaetal2004} observed, indeed, that due to shielding effects high density lines are narrower and therefore colder than low density absorbers.
Whereas \citet{rauchetal2005} found from an investigation of the velocity structure of Ly$\alpha$ forest clouds in the spectra of multiple QSO sight lines that the large scale motions in the IGM are dominated by the Hubble flow.
According to \citet{rauchetal2005} some absorbers undergo early stages of gravitational collapse but the majority of absorbers apparently arise from expanding filaments accelerated by the infall into more massive structures.

In a study of the reionization of helium \citet{gleseretal2005} find that the temperature-density relation may trace the stages of \ion{He}{ii} reionization.
With ongoing reionization the slope of the temperature-density relation gets steeper and a significant scatter is introduced.
At $z \sim 2 -3$ the scatter affects in particular the low density regime.
When helium reionization is completed the slope of the power law flattens again and the amount of scatter decreases since adiabatic cooling dominates over photoionization heating.
Similar results have been obtained by \citet{boltonetal2004} who examined the effects of radiative transfer on the temperature-density relation.
Indeed, the mean temperature of the IGM peaks at $z \sim 3$ \citep{schayeetal2000, theunsetal2002, theunsetal2002c} which is thought to be due to additional heating because of \ion{He}{ii} reionization.

From the analysis of the resolved \ion{He}{ii} Ly$\alpha$ forest in the spectra  of the quasars HE~2347-4342 and HS~1700+6416 a correlation between the density of the absorbers and the hardness of the ionizing radiation (represented by the parameter $\eta$ which is the \ion{He}{ii}/\ion{H}{i} column density ratio) has been found in the sense that higher column density absorbers are apparently exposed to a harder radiation field \citep{shulletal2004, zhengetal2004, reimersetal_fuse, fechneretal2006b}.
There are several suggestions for a physical explanation of this correlation like self-shielding and shadowing effects \citep{maselliferrara2005, tittleymeiksin2006} and the spatial distribution of QSOs as the sources of hard radiation \citep{boltonetal2006}.
However, \citet{fechnerreimers2006} suggested that part of the correlation might be due to the unjustified assumption that the line widths of the absorption features are dominated by purely turbulent broadening.
If this assumption is incorrect, a correlation between the $\eta$
value and the absorber density will be introduced artificially.
In order to reproduce a given equivalent width of an absorption line the column density will be overestimated if the line width is underestimated.
This effect is in particular pronounced for strong absorption features
with high column densities that are beyond the linear part of the
curve of growth and thus introduce an apparent correlation between
$\eta$ and \ion{H}{i} column density. 

Measurements of observed features indicate that the lines are indeed turbulently broadened \citep{zhengetal2004}.
However, this result is based on a limited statistical sample of 8 absorbers.
\citet{liuetal2006} claimed to confirm this finding using hydrodynamic simulations.
On the other hand, \citet{fechnerreimers2006} showed that statistically, lines with apparent low $\eta$ values are often dominated by thermal broadening.
In order to solve this apparent conflict further investigation is needed.

In this work we introduce a simple approach to consider the temperature of the IGM when analyzing the \ion{He}{ii} Ly$\alpha$ forest.
This provides, at least in principle, an independent method to measure the temperature-density relation (often referred to as equation of state of the intergalactic medium).
Since HE~2347-4342 and HS~1700+6416 are the only objects where high resolution FUSE spectra are available \citep{krissetal2001, shulletal2004, zhengetal2004, fechneretal2006b}, we apply our method to these sight lines.
However, as we will see, the data quality ($S/N \sim 5$) is insufficient to derive tight constraints for the temperature-density relation.

After shortly reviewing several equations important in the context of the temperature-density relation in Section \ref{eos_section}, Section \ref{method} presents the methodical approach to derive the parameters of the temperature-density relation and $\eta$ from observations of the \ion{He}{ii} Ly$\alpha$ forest in comparison to \ion{H}{i}.
This method is tested using artificial data in Section \ref{simulations} and applied to the observations in Section \ref{observations}.
In Section \ref{directestimate} we introduce an alternative approach to estimate the temperature for the individual absorption lines more directly and discuss the effect on the inferred $\eta$ values.

\section{The thermal state of the IGM}
\label{eos_section}

\citet{huignedin1997} have shown that there is a well-defined power law relation between temperature and density, often referred to as equation of state for the low density IGM ($\delta \lesssim 10$, i.e.\ $N_{\ion{H}{i}} \lesssim 10^{15}\,\mathrm{cm}^{-2}$).
\begin{equation}\label{hg97}
  T = T_0 \left(1+\delta \right)^{\gamma -1}\,\mbox{,}
\end{equation}
where $\delta = \frac{\rho}{\bar\rho} - 1$ is the gas overdensity and $T_0$ the temperature at the mean density.
This equation of state can be converted into a power law relation between the cut-off Doppler parameter $b_{\mathrm{min}}$ and the \ion{H}{i} column density using cosmological simulations \citep{schayeetal1999}.
Therefore, estimates of the parameters $T_0$ and $\gamma$ are usually based on measuring the cut-off of the $b(N_{\ion{H}{i}})$ distribution.
The combination of \ion{H}{i} and \ion{He}{ii} data would provide, at least in principle, an independent estimate of the thermal state of the IGM.
In order to relate the equation of state to the observed \ion{H}{i} and \ion{He}{ii} spectra it is necessary to connect the \ion{H}{i} column density $N_{\ion{H}{i}}$ and the temperature of the absorbing material.

Assuming that the almost fully ionized gas is in photoionization equilibrium, the number density of neutral hydrogen can be written as
\begin{eqnarray}\label{nhi}
  n_{\ion{H}{i}} & \approx & 1.41\cdot10^{-14}\,\mathrm{cm}^{-3}\,(1+\delta)^2\,(1+z)^6\,\left(\frac{T}{10^4\,\mathrm{K}}\right)^{-0.755} \nonumber \\
   & & \times \left(\frac{\Gamma_{\ion{H}{i}}}{10^{-12}\,\mathrm{s}^{-1}}\right)^{-1}\,\left(\frac{\Omega_{\mathrm{b}}h^2}{0.02}\right)^2\,\mbox{,}
\end{eqnarray}
with the photoionization rate of hydrogen $\Gamma_{\ion{H}{i}}$.
In order to derive this relation we have used
\begin{equation}\label{nh}
  n_{\element{H}} = \frac{3\,\Omega_{\mathrm{b}}\,H_0^2}{8\,\pi\,G\,m_{\element{H}}}\,(1-Y)\,(1+z)^3\,(1+\delta)\,\mbox{,}
\end{equation}
with the primordial mass fraction of helium $Y \approx 0.244$ \citep{burlesetal2001}.

Since $N_{\ion{H}{i}} = n_{\ion{H}{i}}\cdot l$, Eq.\ (\ref{nhi}) leads to a relation between \ion{H}{i} column density and overdensity $\delta$.
However, the length $l$ of the absorbers has to be known.
\citet{bryanmachacek2000} choose a typical size of absorbers from simulations scaling with $l \propto (1+z)^{-1}$.
\citet{schaye2001} argued that the characteristic size of an absorber along any line of sight will generally be in order of the local Jeans length
\begin{equation}\label{jeanslength}
  L_{\mathrm{J}} \approx \sqrt{3.59\cdot10^{42}\,\mathrm{cm}^{-1}\,\left(\frac{T}{10^4\,\mathrm{K}}\right)\left(\frac{f_{g}}{0.16}\right)\,n_{\element{H}}^{-1}}\,\mbox{,}
\end{equation}
with the fraction $f_g$ of mass in the gas which should be close to the universal value $f_g \approx \Omega_{\mathrm{b}}/\Omega_{\mathrm{m}} \sim 0.16$ in the low density Ly$\alpha$ forest.
Combining Eqs.\ (\ref{nhi}), (\ref{nh}), and (\ref{jeanslength}) leads to an analytic expression for the relation between the observable column density of neutral hydrogen $N_{\ion{H}{i}}$ and the corresponding overdensity \citep[see also][]{schaye2001}
\begin{eqnarray}\label{s01}
  N_{\ion{H}{i}} & \approx & 6.50\cdot10^{10}\mathrm{cm}^{-2}\,(1+\delta)^{1.5}\,(1+z)^{4.5}\left(\frac{T}{10^4\,\mathrm{K}}\right)^{-0.255} \nonumber\\
  & &\times\,\left(\frac{\Gamma_{\ion{H}{i}}}{10^{-12}\,\mathrm{s}^{-1}}\right)^{-1}\,\left(\frac{\Omega_{\mathrm{b}}h^2}{0.02}\right)^{1.5}\left(\frac{f_g}{0.16}\right)^{0.5}\,\mbox{.}
\end{eqnarray}
The latter three parameters are in the order of unity.
Then, inserting the equation of state yields
\begin{eqnarray}\label{logNHI_logT}
  \log N_{\ion{H}{i}} & \approx & 11.833 + 4.5\cdot\log(1+z) - \frac{1.5}{\gamma -1}\log T_0 \nonumber \\
  & & + \left(\frac{1.5}{\gamma -1}-0.255\right)\cdot\log T\,\mbox{,}
\end{eqnarray}
where $N_{\ion{H}{i}}$ is measured in $\mathrm{cm}^{-2}$ and the temperatures are given in K.
Thus, we have found a relation between the \ion{H}{i} column density and the temperature of an absorber which only depends on the redshift and the parameters of the equation of state $\gamma$ and $\log T_0$.
However, its validity is limited to gas with low and moderate
densities. 
At higher densities ($N_{\ion{H}{i}} \gtrsim 10^{15}\,\mathrm{cm}^{-2}$) radiative cooling dominates over adiabatic
cooling due to the Hubble expansion, leading to lower temperatures.
Furthermore, a possible scatter in the temperature-density relation due to helium reionization \citep{gleseretal2005} is neglected.

\section{Method}\label{method}

Three different analysis methods have been applied to the observations of the \ion{He}{ii} Ly$\alpha$ forest so far: line profile fitting \citep{krissetal2001, zhengetal2004, fechneretal2006b}, an apparent optical depth method \citep{shulletal2004, fechneretal2006b} and spectrum fitting \citep{fechnerreimers2006}.
The latter fits the high-quality optical \ion{H}{i} spectrum directly to the \ion{He}{ii} data, and allows to estimate the length scales of $\eta$ variations.
Among these three methods only profile fitting provides the possibility of taking into account different line widths of \ion{H}{i} and \ion{He}{ii} as would be introduced by thermal line broadening.
In order to preserve the advantage of the spectrum fitting method to estimate scales on which the ionizing background fluctuates, we combine the traditional profile fit analysis and the spectrum fitting method.

The \ion{H}{i} Ly$\alpha$ forest is fitted with Doppler profiles.
From the resulting line list an artificial \ion{He}{ii} spectrum is computed and convolved with an appropriate instrumental profile.
The artificial spectrum can then be fitted to the FUSE data by scaling the \ion{H}{i} column density.
Furthermore, the line width of the \ion{He}{ii} features can be addressed via the Doppler parameter $b_{\ion{He}{ii}}$ which could lead in principle to an independent estimate of the temperature of the IGM and the parameters of the equation of state.
As we will see below, fitting the temperature is sensitive to the noise level of the \ion{He}{ii} spectrum. 
The quality of present day \ion{He}{ii} data from FUSE is insufficient to derive well constrained temperatures and scales of variation for $\eta$ simultaneously.
Consequently, we will a priori select redshift ranges where $\eta$ is assumed to be constant.

Fitting line profiles introduces further uncertainties.
As demonstrated by \citet{fechneretal2006b}, a scatter of $10 - 15\,\%$ in $\eta$ is expected even if the underlying value is constant.
This is due to the fact that any line profile model recovers the observed data only imperfectly.
Additionally, the high noise level of the FUSE data introduces further uncertainties \citep[see also][]{liuetal2006}.
Since these problems are inevitable, we expect that our results can be interpreted statistically but conclusions for single absorbers should be considered very carefully.

The modified spectrum fitting method works as follows:
First, the \ion{H}{i} data is fitted and a line list with the values of redshift $z$, column density $\log N_{\ion{H}{i}}$, and Doppler parameter $b_{\ion{H}{i}}$ is prepared.
Then values of the parameters $\log T_0$ and $\gamma$ are selected.
The temperature $T$ of each single line is computed according to Eq.\ (\ref{logNHI_logT}).
The Doppler parameter of \ion{He}{ii} is then given by
\begin{equation}\label{bheii}
  b_{\ion{He}{ii}} = \sqrt{b_{\ion{H}{i}}^{2} - 2\,k\,T\cdot\left(\frac{1}{m_{\element{H}}} - \frac{1}{m_{\element{He}}}\right)}\,\mbox{,}
\end{equation}
where $k$ is Boltzmann's constant and $m_{\element{H}}$ and $m_{\element{He}}$ are the masses of the hydrogen or helium atom, respectively.
If the temperature is very high, the term beneath the square-root in Eq.\ (\ref{bheii}) may become negative.
In this case the Doppler parameter of \ion{H}{i} is assumed to be completely thermal and the temperature is set to $T = b_{\ion{H}{i}}^2\cdot m_{\element{H}}/(2\,k)$.
From the adjusted line parameters an artificial \ion{He}{ii} spectrum is generated and fitted to the data using a $\chi^2$-procedure where $\log\eta$ is varied \citep[as described in][]{fechnerreimers2006}.
This means, we assume $\eta$ to be constant in the considered redshift range.
This procedure is repeated for several values of $\log T_0$ and $\gamma$.
We explore the parameter space $3.50 \le \log T_0 \le 4.50$ and $0.70 \le \gamma \le 1.80$ where realistic results are expected.
By comparing the resulting $\chi^2$-values for each combination of $\log T_0$ and $\gamma$ the best fit parameters can be found.
In total three free parameters ($\log\eta$, $\log T_0$, and $\gamma$) are estimated for a given redshift interval.

Eq.\ (\ref{bheii}) assumes that the thermal and turbulent Doppler
parameters both lead to a Gaussian profile and combine to the total
Doppler parameter as $b^{2} = b_{\mathrm{therm}}^{2} +
b_{\mathrm{turb}}^{2}$.
Since the turbulent part of the $b$-parameter summarizes all kinds of
velocities except thermal, the broadening profile is not necessarily
Gaussian.
In particular, due to the atomic parameters of helium, absorbers with
moderate \ion{He}{ii} column densities probe the low density
Ly$\alpha$ forest and are probably dominated by the differential Hubble flow rather than microscopic turbulence.
Thus, Eq.\ (\ref{bheii}) represents the most simple approach to
combine thermal and turbulent line broadening.
This should be sufficient for a first analysis.

\section{Application to artificial data}\label{simulations}

In order to explore the potential of the introduced approach we
examine artificial data which is created on the basis of the
statistical properties of the Ly$\alpha$ forest.
The column density distribution function with $\beta = 1.5$ is adopted
from \citet{kirkmantytler1997}.
Values in the range $11.0 \le \log N_{\ion{H}{i}} \le 18.0$ are
simulated.
The Doppler parameter distribution is described by a truncated
Gaussian ($b_{\mathrm{min}} = 10\,\mathrm{km\,s}^{-1}$) centered at
$27\,\mathrm{km\,s}^{-1}$ with a with of $\sigma_{b} = 8.75
\,\mathrm{km\,s}^{-1}$ in agreement with \citet{huetal1995}.
In addition the temperature is linked to the \ion{H}{i} column density according to Eq.\ (\ref{logNHI_logT}).
The parameters of the temperature-density relation are set to the realistic values $\log T_0 = 4.3$ and $\gamma = 1.3$ \citep[e.g.][]{schayeetal2000}.
Furthermore, we chose $\log\eta = 1.903$ \citep[i.e.\ $\eta = 80$; e.g.][]{krissetal2001}.
The spectra cover the redshift range $2.56 < z < 2.92$ with $R =
40\,000$ in the \ion{H}{i} and $R = 15\,000$ in the \ion{He}{ii} Ly$\alpha$ forest.
In order to study the importance of the data quality several signal-to-noise ratios are chosen for the artificial \ion{He}{ii} spectrum ($S/N = 5$, $10$, $20$, $30$, $50$, $100$).
The $S/N$ for the \ion{H}{i} Ly$\alpha$ forest is set to $100$ which is comparable to the noise level of the best optical spectra available today and is sufficient to derive a reasonable line list from the fit.

\begin{figure}
  \centering
  \resizebox{\hsize}{!}{\includegraphics[bb=65 35 355 465,clip=,angle=-90]{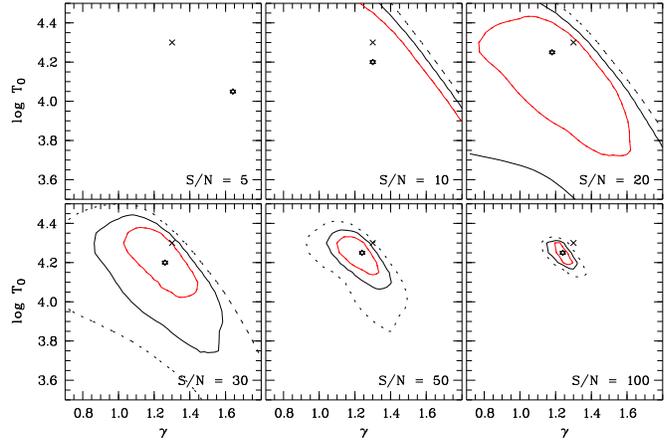}}
  \caption{Contour plots for the parameters of the temperature-density relation derived from the simulated data.
The crosses represent the position of the true values of $\gamma = 1.3$ and $\log T_0 = 4.3$ while the stars indicate the estimated minimum.
The outer contours give the joint $1\,\sigma$ (solid line) and $2\,\sigma$ (dotted line) confidence levels, respectively.
The inner contour represents the $1\,\sigma$ confidence level for one free parameter, i.e.\ the projection of the inner contour to the axes leads to the $1\,\sigma$ errors of $\log T_0$ and $\gamma$, respectively.
The signal-to-noise ratio of the artificial \ion{He}{ii} spectra is indicated in each panel.
The confidence levels in case of $S/N = 5$ (upper left panel) are outside the presented parameter range.
  }
  \label{contours_sim}
\end{figure}
 
\begin{figure}
  \centering
  \resizebox{\hsize}{!}{\includegraphics[bb=80 35 280 665,clip=,angle=-90]{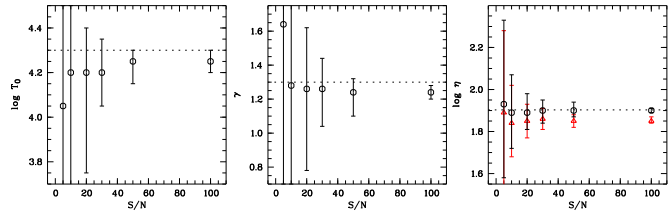}}
  \caption{Results for the fit of the artificial data.
The panels show the derived values for $\log T_0$, $\gamma$, and $\log\eta$ (from the left to the right).
The error bars correspond to the inner $1\,\sigma$ confidence contours of Fig.\ \ref{contours_sim}.
The dotted lines indicate the true values.
For comparison the $\eta$ values estimated under the assumption of pure turbulent line broadening are presented as triangles in the right panel.
  }
  \label{results_sim}
\end{figure}

The \ion{H}{i} Ly$\alpha$ forest is fitted with Doppler profiles once and afterwards all artificial \ion{He}{ii} spectra are fitted on the basis of the derived line list according to the method presented in Section \ref{method}.
Fig.\ \ref{contours_sim} shows the $1\,\sigma$ and $2\,\sigma$ confidence contours for the parameters $\log T_0$ and $\gamma$ of the temperature-density relation (Eq.\ \ref{hg97}).
The point of the best fit is marked by a star, while the true underlying value is indicated by a cross.
As expected, the results are better constrained if the $S/N$ of the \ion{He}{ii} spectrum is increased.
In case of $S/N = 5$ a reasonable estimate cannot be derived since the variation of the absorber's temperature affects the wings of the line profiles only marginally and due to the high noise level very similar $\chi^2$-values are produced.
In order to derive reasonable constraints for the temperature-density relation a signal-to-noise ratio of at least $S/N \sim 20\dots 30$ is required for the \ion{He}{ii} data.
The results are summarized in Fig.\ \ref{results_sim} which presents the best fit values and their $1\,\sigma$ uncertainties in comparison to the true underlying values.
It becomes clear that also for $S/N = 100$ the derived $\log T_0$ and $\gamma$ are underestimated.
The reason for this systematic effect is the imperfectness of the adopted line list (see below).
However, the inferred $\eta$ values are in good agreement with the underlying $\eta = 80$.

For comparison we also perform a fit assuming pure turbulent line widths, i.e.\ $b_{\ion{He}{ii}} = b_{\ion{H}{i}}$. 
The resulting $\eta$ values are presented as triangles in the right panel of Fig. \ref{results_sim}.
Obviously, the $\eta$ value is underestimated if pure turbulent line broadening is assumed.
This finding is essentially independent of the signal-to-noise ratio of the \ion{He}{ii} and also of the modifications of the line list (see below).
On average the pure turbulent $\eta$ value is $\sim
0.05\,\mathrm{dex}$ lower than the $\eta$ derived when considering a temperature-density relation and thus about $0.05\,\mathrm{dex}$ lower than the underlying value.
This is still consistent within the $1\,\sigma$ level in case of low $S/N$ but the underestimation will be significant as the error bars decrease for higher quality data ($S/N \gtrsim 50$).

\begin{figure}
  \centering
  \resizebox{\hsize}{!}{\includegraphics[bb=395 35 550 465,clip=,angle=-90]{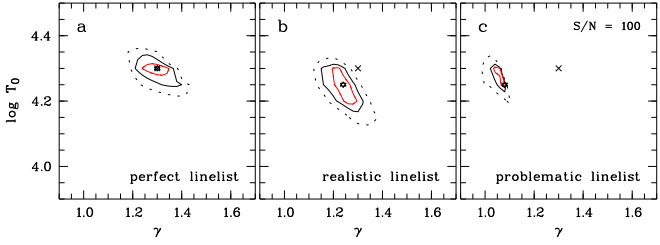}}
  \caption{Contour plots for the parameters of the temperature-density relation derived from the artificial data using three different \ion{H}{i} line lists for the \ion{He}{ii} fitting (see text).
The crosses represent the position of the true values of $\gamma = 1.3$ and $\log T_0 = 4.3$ while the stars indicate the estimated minimum.
Contour levels are given as in Fig.\ \ref{contours_sim}.
  }
  \label{contours_linelists}
\end{figure}
 
As noted above, the best fit depends on the quality of the adopted line list.
This is illustrated in Fig.\ \ref{contours_linelists}.
We performed fits to the artificial \ion{He}{ii} spectrum with $S/N = 100$ based on three different \ion{H}{i} line lists.
The ``perfect'' line list includes all absorption lines that have been generated to compute the artificial spectra.
In this case the correct parameters are, of course, reproducible (Fig.\ \ref{contours_linelists}a).
We find $\log T_0 = 4.30\pm0.02$ and $\gamma = 1.30^{+0.04}_{-0.06}$.
For a ``reasonable'' line list based on a thorough fit of the \ion{H}{i} spectrum we obtain $\log T_0 = 4.25\pm0.05$ and $\gamma = 1.24\pm0.04$ (Fig.\ \ref{contours_linelists}b).
In particular $\gamma$ is not reproduced correctly although the estimated value has apparently small error bars.
In this case the imperfectness of the line list introduces additional systematic uncertainties.

\begin{figure}
  \centering
  \resizebox{\hsize}{!}{\includegraphics[bb=35 405 315 725,clip=]{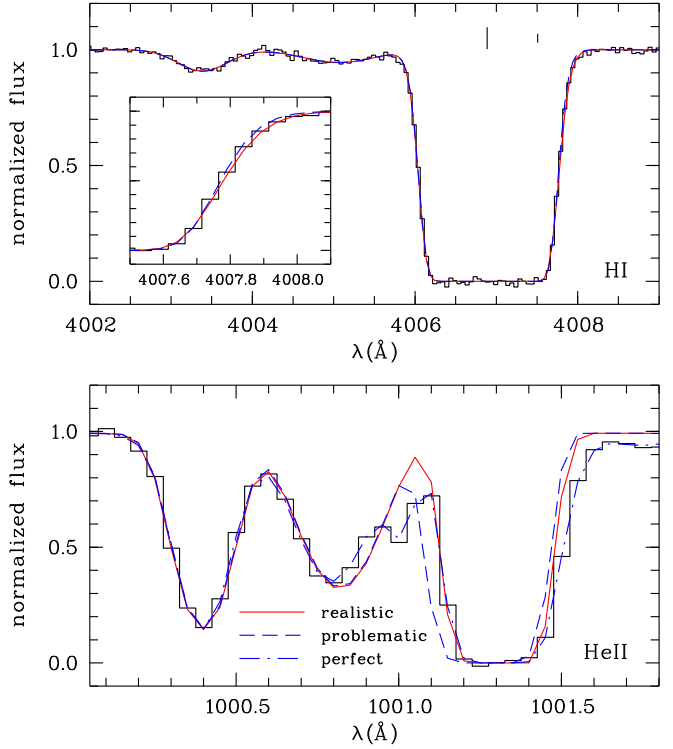}}
  \caption{Illustration of the importance of the \ion{H}{i} fit.
The upper panel shows a portion of the artificial \ion{H}{i} data (histogram-like) and two different fits where the saturated feature at $4007\,\mathrm{\AA}$ is modelled with one (dashed line) or two (solid line) Doppler components, respectively.
The inset in the upper panel illustrates the only marginal difference between both fits.
The lower panel shows the artificial \ion{He}{ii} data (histogram-like) and the best fits in case of the one component approach (dashed line; $\log T_0 \approx 4.25$, $\gamma \approx 1.08$) or two component fit (solid line; $\log T_0 \approx 4.25$, $\gamma \approx 1.24$), respectively.
The dot-dashed line represents the best fit using the perfect line list ($\log T_0 = 4.30$, $\gamma = 1.30$).
  }
  \label{linelist_profiles}
\end{figure}

Furthermore, we examine a realistic but ``problematic'' case (Fig.\ \ref{contours_linelists}c).
The difference between the realistic and the problematic line list is illustrated in Fig.\ \ref{linelist_profiles}. 
A broad, saturated \ion{H}{i} feature has been fitted with two Doppler profiles (realistic case) or with one component only (problematic case).
The upper panel of Fig.\ \ref{linelist_profiles} shows only very slight discrepancies between the resulting profiles.
However, there are more prominent differences in the \ion{He}{ii} fit (lower panel of Fig.\ \ref{linelist_profiles}).
As a consequence of the absence of a second component the feature has to be modeled  by a broader profile to achieve a reasonable fit.
Therefore, the temperature $T$ has to be lower according to Eq.\ (\ref{bheii}).
Because of the given \ion{H}{i} column density of the line $\gamma$ has to decrease (see Eq.\ \ref{logNHI_logT}).
Indeed, we find $\log T_0 = 4.25^{+0.05}_{-0.01}$ and $\gamma = 1.08^{+0.01}_{-0.04}$ (Fig.\ \ref{contours_linelists}c).
Additionally, a weak \ion{H}{i} line ($\log N_{\ion{H}{i}} = 11.7$) leading to a more prominent \ion{He}{ii} feature ($\log N_{\ion{He}{ii}} = 13.4$) at $1001.0\,\mathrm{\AA}$ is missed in both models introducing further deviations from the optimum model.

However, in case of observed spectra the best strategy to produce the \ion{H}{i} line list would also take into account higher order features of the Lyman series.
Thus, the estimation of the line parameters of a saturated absorption complex would be more accurate than in this simple simulated case.
Due to blending or incomplete spectral coverage a few features with ambiguous line parameters might still be left unsettled even for observed spectra.
But the existence of unidentified, weak \ion{H}{i} features below the detection limit can never be excluded.
For this reason profile fit based analyses of the \ion{He}{ii} Ly$\alpha$ forest include \ion{He}{ii} lines without detected \ion{H}{i} counterparts \citep{krissetal2001, zhengetal2004, fechneretal2006b}.
One should keep in mind that even one missing component may affect the results significantly.

\section{Application to observations}\label{observations}

Though the data quality for the two \ion{He}{ii} quasars observed with FUSE, HE~2347-4342 and HS~1700+6416 ($S/N \sim  5$), is well below the required noise level of $S/N \gtrsim 20$, we apply the method presented in Sect.\ \ref{method} to the observed spectra.
Our main intention is to explore whether any conclusions can be inferred even from this noisy data.
In order to avoid additional degrees of freedom the scales to perform a fit are selected a priori.
We select spectral ranges with a size of $\sim 5\,\mathrm{\AA}$, corresponding to $\Delta z \sim 0.016$.
Scales of variation of the UV background in this order of magnitude have been estimated by \citet{fechnerreimers2006}, whereas \citet{shulletal2004} and \citet{boltonetal2006} find much smaller scales in the range $\Delta z \sim 0.001$.
Realistically, we do not expect to be able to derive robust constraints on the parameters of the temperature-density relation.
However, this approach should be sufficient to roughly estimate the effect of partly thermal Doppler parameters on the results. 

The far-UV data of HE~2347-4342 are presented in \citet{zhengetal2004}.
We use the normalization described in \citet{fechnerreimers2006}.
The spectrum is binned to $0.05\,\mathrm{\AA}$ which corresponds to
the resolution of $R = 20\,000$ at $1000\,\mathrm{\AA}$ (and the actual value $\sim 15\,\mathrm{km\,s}^{-1}$) and reduces the noise to a $S/N \sim 5$ level per bin in the whole spectrum.
In order to avoid further uncertainties we exclude the spectral range containing Ly$\beta$ or higher order Lyman series lines.
For the \ion{H}{i} Ly$\alpha$ forest towards HE~2347-4342 we adopt the line list of \citet{fechnerreimers2006} based on a Doppler profile fit of high-quality VLT/UVES spectra ($R \approx 45\,000$, $S/N \approx 100$).
We slightly modify the parameters of the Lyman limit system complex at $z \sim 2.735$ which are now estimated using the Lyman series up to Ly-9.

The FUSE data of HS~1700+6416 are presented in \citet{fechneretal2006b}.
It has a resolution of $R \sim 20\,000$.
The signal-to-noise ratio is $\sim 5$ and slightly better.
The line list for \ion{H}{i} Ly$\alpha$ forest is adopted from the results of \citet{fechneretal2006b} who use high-quality data from Keck/HIRES ($R \approx 38\,500$, $S/N \approx 100$).
Furthermore, we correct for metal lines in the FUSE spectral range as described by \citet{fechnerreimers2006} using the models of \citet{fechneretal2006a}.

\begin{figure}
  \centering
  \resizebox{\hsize}{!}{\includegraphics[bb=45 405 415 755,clip=]{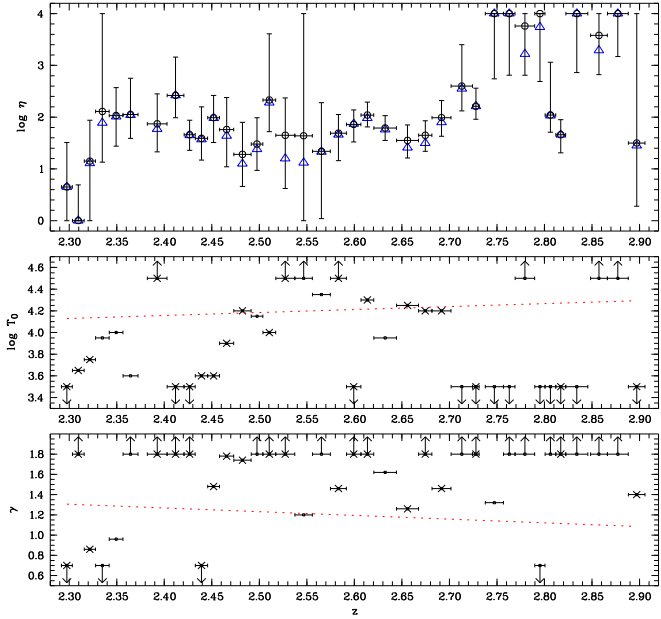}}
  \caption{Results from HE~2347-4342 for $\log\eta$ (open circles; upper panel), $\log T_0$ (middle panel), and $\gamma$ (lower panel).
The triangles in the upper panel present the resulting $\eta$ value assuming pure turbulent broadening.
For clarity error bars are not given for the turbulent case, but are of the same size as those of the temperature-density relation fit.
The $1\,\sigma$ uncertainties of $\log T_0$ and $\gamma$ would comprise the whole presented parameter range.
Fit sections of $\sim 5\,\mathrm{\AA}$ have been selected a priori.
Sections marked with a cross represent redshift intervals without
extremely saturated absorption features.
The dotted lines in the lower panels indicated the expected value interpolated from the measurements of \citet{schayeetal2000}.
  }
  \label{he2347_summary}
\end{figure}

\begin{figure}
  \centering
  \resizebox{\hsize}{!}{\includegraphics[bb=45 405 415 755,clip=]{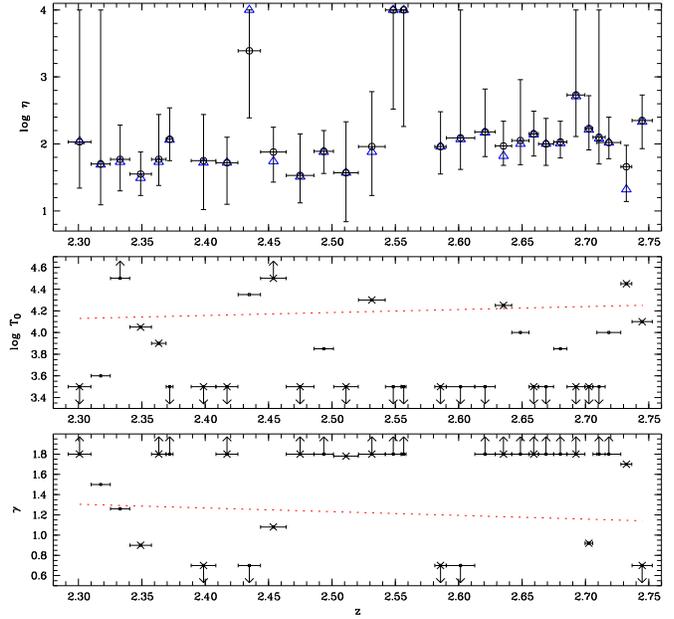}}
  \caption{As Fig.\ \ref {he2347_summary} but for HS~1700+6416.
  }
  \label{hs1700_summary}
\end{figure}

As a compromise between constructing a model as realistic as possible and the caveats due to the limited quality of the data we fit the selected spectral ranges assuming constant values of $\eta$, $\log T_0$, and $\gamma$. 
For comparison we also estimate a fit assuming pure turbulent line widths.
The results are presented in Figs.\ \ref{he2347_summary} (HE~2347-4342) and \ref{hs1700_summary} (HS~1700+6416).
Towards HE~2347-4342 the fit intervals at $z \gtrsim 2.75$ cover the range of patchy \ion{He}{ii} absorption \citep{reimersetal1997} which marks the tail end of the epoch of \ion{He}{ii} reionization \citep[also e.g.][]{krissetal2001, zhengetal2004}.
For these redshift ranges further uncertainties like the incomplete reionization of \ion{He}{ii} \citep[e.g.][]{reimersetal1997, reimersetal2005c} or simply the fact that the FUSE spectrum shows completely saturated, broad absorption troughs, lead to extremely unstable fits which should not be considered further.

\begin{figure}
  \centering
  \resizebox{\hsize}{!}{\includegraphics[bb=35 540 355 770,clip=]{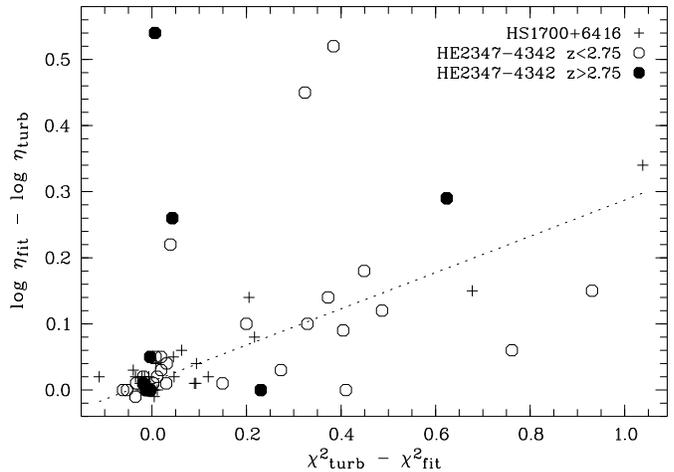}}
  \caption{Correlation between the difference in the quality of the fit, i.e.\ the $\chi^2$, and the difference in the $\eta$ value.
The better the fit taking into account a temperature-density relation in comparison the the fit assuming pure turbulent broadening, the larger is the resulting $\log\eta$ for the best fit.
The dotted line represents a linear fit to the Ly$\alpha$ forest data, i.e.\ HS~1700+6416 and HE~2347-4342 at redshifts $z < 2.75$.
  }
  \label{chi2_correlation}
\end{figure}

Generally, the $\eta$ values in the pure turbulent cases are slightly lower than if the thermal state of the IGM is taken into account, consistent with the results from the artificial data.
On average the turbulent $\eta$ is lower by roughly $0.05\,\mathrm{dex}$ considering the Ly$\alpha$ forest only (i.e.\ $z < 2.75$).
The difference in the $\eta$ value is correlated with the quality of the fit, i.e.\ with the $\chi^2$, which is shown in Fig.\ \ref{chi2_correlation}.
If a thermal component to the line width is considered, the \ion{He}{ii} lines become narrower in comparison to the pure turbulent case. 
In order to produce roughly the same equivalent width the column density of \ion{He}{ii} and thus $\eta$ has to increase.
The correlation presented in Fig.\ \ref{chi2_correlation} means if considering the thermal state of the absorbers significantly improves the fit in a selected \ion{He}{ii} wavelength range, the inferred $\eta$ value will be larger.
And even if the fit is improved only slightly ($\chi^2_{\mathrm{turb}} - \chi^2_{\mathrm{fit}} < 0.1$), the $\eta$ value increases on average by $0.02\,\mathrm{dex}$ (median $0.01\,\mathrm{dex}$).
However, the deviations are still within the $1\,\sigma$ uncertainties of the estimate (see also the upper panels of Figs.\ \ref{he2347_summary} and \ref{hs1700_summary}).

The parameters of the temperature-density relation estimated by the fits are rather unconstrained.
As described in Sect.\ \ref{method} we explored the parameter space $3.50 \le \log T_0 \le 4.50$ and $0.70 \le \gamma \le 1.80$.
Within these limits all values produce models fitting to the data on a $1\,\sigma$ level.
Therefore, no error bars are presented in Figs.\ \ref{he2347_summary} and \ref{hs1700_summary}.
Nevertheless, some tendencies are noteworthy.

The vast majority of the fit intervals exhibit an extreme value for $\gamma$ ($\le 0.7$ or $\ge 1.8$) and/or $\log T_0$ ($\le 3.5$ or $\ge 4.5$).
Excluding the redshift range with patchy \ion{He}{ii} absorption ($z \gtrsim 2.75$) from further discussion, 81\,\% of the fitting sections yield extreme values of the temperature-density relation.
Among these cases 34\,\% lead to a temperature-density relation with $\gamma \ge 1.8$ and $\log T_0 \le 3.5$.

\begin{figure}
  \centering
  \resizebox{\hsize}{!}{\includegraphics[bb=40 475 455 740,clip=]{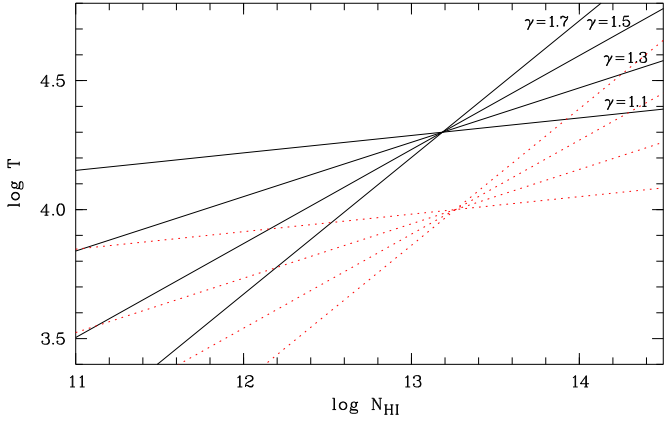}}
  \caption{Relation between \ion{H}{i} column density $\log N_{\ion{H}{i}}$ and the absorber's temperature $\log T$ according to Eq.\ (\ref{logNHI_logT}).
The relations are shown for $\log T_0 = 4.3$ (solid lines) and $\log T_0 = 4.0$ (dotted lines) and $\gamma = 1.7$, $1.5$, $1.3$, and $1.1$. 
  }
  \label{logN_logT_examples}
\end{figure}

In order to give an interpretation of the temperature-density relation in this case some examples are illustrated in Fig.\ \ref{logN_logT_examples}.
A large value of $\gamma$ leads to a steep relation between column density and temperature.
Consequently, the temperature of high column density absorbers is much higher than the temperature of low column density absorbers.
The parameter $\log T_0$ rules the absolute values of the temperature.
The lower $\log T_0$, the lower is the general temperature level.
Finding $\gamma \ge 1.8$ means that the line widths $b_{\ion{He}{ii}}$ of high column density absorbers are significantly narrower than $b_{\ion{H}{i}}$.
Therefore thermal line broadening is favored.
This idea may support the conclusions of \citet{fechnerreimers2006} who suggest that thermal line broadening might be important for high column density absorbers.

For about 9\,\% of the extreme cases we find $\gamma \le 0.70$ and $\log T_0 \le 3.50$, the lowest values in the parameter range explored.
At this point the lowest temperatures are produced for high column density absorbers since the slope ($\gamma - 1$) of the temperature-density relation gets negative and $\log T_0 = 3.5$ sets the absolute scale to low temperatures.
Thus, thermal broadening might play a minor role for the line widths of the \ion{H}{i} and \ion{He}{ii} features.
For these fit sections the $\eta$ values estimated from the pure turbulent and the thermal state model are nearly identical supporting the idea that turbulent line broadening is dominating in these redshift ranges.
At $z \sim 2.43$ towards HS~1700+6416 both derived $\eta$ values are
heavily overproduced due to problems with two saturated, broad
absorption features that dominate the selected spectral portion.

The results are expected to be more stable if regions with saturated
or extremely weak \ion{H}{i} are excluded.
The data points marked with a cross in the lower panels of Figs.\
\ref{he2347_summary} and \ref{hs1700_summary} represent the sections
free from strongly saturated features.
However, the results presented above for the whole spectra is also
valid when considering only the selected portion of the data.
Differences should be expected for data of higher quality.

\begin{figure}
  \centering
  \resizebox{\hsize}{!}{\includegraphics[bb=42 600 320 753,clip=]{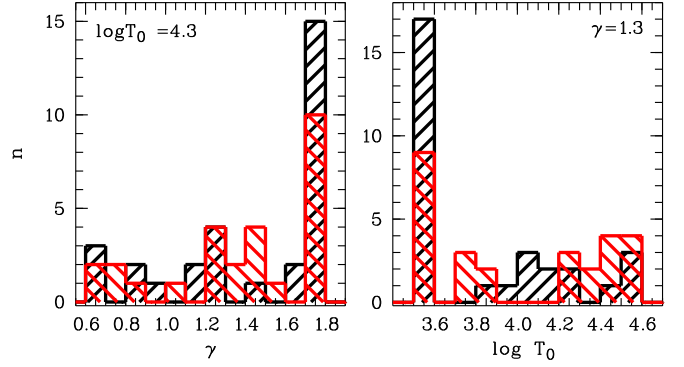}}
  \caption{Distribution of the parameters of the temperature-density
  relation estimated individually for the sight lines towards
  HS~1700+6416 (black) and HE~2347-4342 (light; $z < 2.75$).
  The value of $\log T_0 = 4.3$ has been fixed when estimating
  $\gamma$ (left panel), while $\gamma = 1.3$ when $\log T_0$ has been
  optimized (right panel).
  }
  \label{par_fixed}
\end{figure}

Furthermore, we test to constrain the parameters of the
temperature-density relation by fixing one parameter at a reasonable value and
varying the other one and $\log\eta$ simultaneously.
The estimated distributions of $\log\eta$ are very similar to those
presented in Figs.\ \ref{he2347_summary} and \ref{hs1700_summary}.
The rms-deviation from the distributions shown is $\sim 0.15$ or even smaller.
Fixing $\log T_0 = 4.3$ and varying $\gamma$ still leads to
inconclusive estimates of $\gamma$ for both lines of sight.
The derived distributions are presented in the left panel of Fig.\ \ref{par_fixed}.
The majority of redshift bins prefers an extreme value $\gamma \ge
1.8$ like in the standard procedure.
Furthermore, the distribution might peak at $\gamma \sim 1.2$, which
would be consistent with independent estimates from the literature
(see below).
However, the detection is insignificant for the present data.
Similar results are found if $\gamma = 1.3$ is fixed.
The estimated distributions of $\log T_0$ are shown in the right panel
of Fig.\ \ref{par_fixed}.
In this case low values of $\log T_0$ are preferred like in Figs.\ \ref{he2347_summary} and \ref{hs1700_summary}.
There might be a slight tendency for $\log T_0 \sim 4.4$ considering
the HE~2347-4342 line of sight, while towards HS~1700+6416 $\log T_0
\sim 4.1$.
However, these results are rather inconclusive.
In order to give more significant constraints, \ion{He}{ii} data of better
quality is needed or at least a more reliable statistics based on a
larger sample.

In the literature \citet{bryanmachacek2000}, \citet{ricottietal2000}, \citet{schayeetal2000}, and \citet{mcdonaldetal2001} measured the parameters of the equation of state in the considered redshift range ($2.3 < z < 2.9$) estimating the lower cut-off of the $b(N_{\ion{H}{i}})$ distribution.
Their resulting values are in the range $0.8 \lesssim \gamma \lesssim 1.7$ and $4.05 \lesssim \log T_0 \lesssim 4.45$ with a strong redshift dependence.
Because of the just completed epoch of \ion{He}{ii} reionization $\gamma$ increases steeply towards lower redshifts while $\log T_0$ decreases with $z$.
Fitting a straight line to the results of \citet{schayeetal2000} in the redshift range $1.8 < z < 3.2$ we find $\gamma = -(0.37 \pm 0.16)\cdot z + (2.15 \pm 0.42)$ and $\log T_0 = (0.27 \pm 0.06)\cdot z +(3.50 \pm 0.15)$, respectively.
The results of \citet{schayeetal2000} are selected since they cover a large redshift range including the portion we are interested in.
The other authors use slightly different methods and provide fewer data points \citep[difference between the methods and systematic offsets in the results are discussed in][]{bryanmachacek2000, ricottietal2000, mcdonaldetal2001}.
For comparison the fit to the \citet{schayeetal2000} data is indicated in Figs.\ \ref{he2347_summary} and \ref{hs1700_summary} as dotted lines.
It is obvious that our results scatter over the range given by the literature and beyond.
Furthermore, there is no significant concentration of measurements towards the \citet{schayeetal2000} interpolation.
Therefore, a detailed comparison is meaningless.
As demonstrated in Section \ref{simulations} UV data of higher quality ($S/N \gtrsim 20$) would be needed to derive tighter constraints for the temperature-density relation using the \ion{He}{ii} Ly$\alpha$ forest.

In addition, recent indications for a small value of $\gamma \approx 1.1$ found by \citet{dodoricoetal2006} investigating the clustering properties in a sample of close QSO pairs, or even hints of an inverses temperature-density relation ($\gamma \approx 0.5$) which would improve a fit of the observed Ly$\alpha$ flux probability distribution function of a sample of 63 quasars reported by \citet{beckeretal2006} cannot be addressed here due to the limited data quality of the \ion{He}{ii} spectra.

\section{Direct temperature estimate}\label{directestimate}

Even though we showed in the previous Sections that an estimate of the IGM temperature requires better UV data, we also follow an alternative approach to estimate the temperatures of the absorbers more directly.
An advantage of this alternative procedure is that the scale of the UV background variations can be estimated.
Therefore, the method described in Section \ref{method} is adjusted.
Instead of the temperature-density relation the temperature of each line itself is fitted by varying the Doppler parameter of \ion{He}{ii} according to Eq.\ (\ref{bheii}).
We start with the pure turbulent case $b_{\ion{He}{ii}} = b_{\ion{H}{i}}$ and then change the thermal component.
Furthermore, a scale estimation as described in \citet{fechnerreimers2006} is implemented.

A fit is computed for a given size of an interval.
Then the bin size is increased until the $\chi^2$ of the best fit gets worse again.
While the spectrum fit method uses the observed optical data directly and may therefore facilitate a pixelwise increment of the fit interval, the approach presented here is based on a superposition of Doppler profiles.
Therefore it has to be ensured that all components possibly contributing at a certain wavelength are taken into account.
Experimenting with several strategies in test calculations we choose a starting interval of $0.5\,\mathrm{\AA}$ in \ion{He}{ii} which is increased by 3 pixels, i.e.\ $0.15\,\mathrm{\AA}$, per iteration.
Furthermore, if the enlargement of the fit interval leads to a larger $\chi^2$, the interval is increased (up to 5 times) to test whether a better fit can be obtained.

\begin{figure}
  \centering
  \resizebox{\hsize}{!}{\includegraphics[bb=35 450 550 770,clip=]{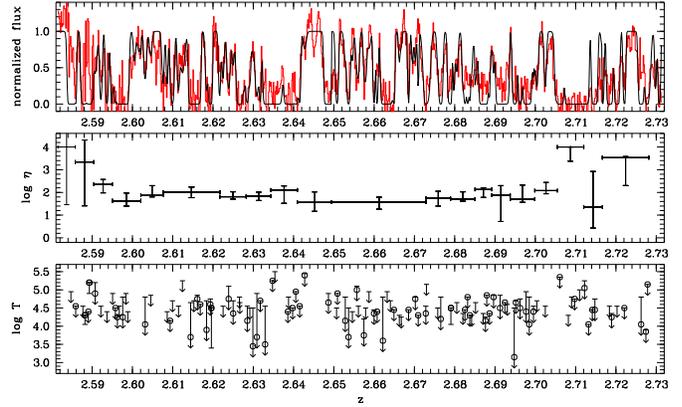}}
  \caption{Best fit towards HE~2347-4342 in the redshift range $2.58 \le z \le 2.73$ (upper panel).
Fitted parameters are the temperatures of the absorbers (lower panel), the \ion{He}{ii}/\ion{H}{i} column density ratio $\eta$ (middle panel), and the scale of $\eta$ variation.
The resulting temperature are presented as open circles in the lower panel.
However, most of them have large error bars and are actually upper limits. 
No temperature is given if turbulent line broadening is preferred (see also text).
At $z \sim 2.69$ two saturated \ion{H}{i} lines appear to be stronger
than the corresponding \ion{He}{ii} features in the line center leading to a poor fit in this range.
  }
  \label{he2347_tempfit}
\end{figure}

\begin{figure}
  \centering
  \resizebox{\hsize}{!}{\includegraphics[bb=35 450 550 770,clip=]{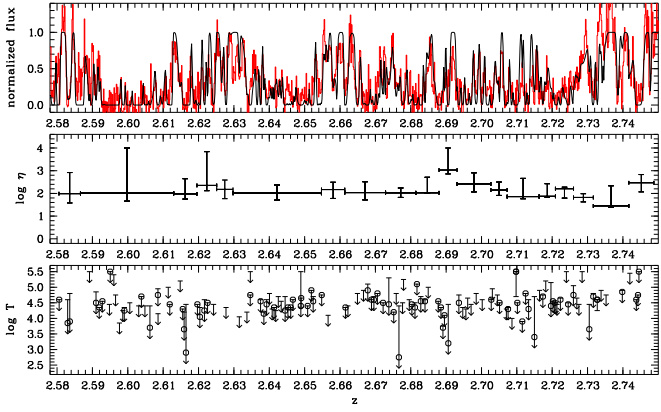}}
  \caption{Best fit and resulting parameters towards HS~1700+6416 in the redshift range $2.58 < z < 2.75$; lines and symbols as in Fig.\ \ref{he2347_tempfit}.
  }
  \label{hs1700_tempfit}
\end{figure}

The best fit for the redshift range $2.58 < z < 2.74$ towards HE~2347-4342 is presented in Fig.\ \ref{he2347_tempfit}.
The corresponding fit of the spectrum of HS~1700+6416 is shown in Fig.\ \ref{hs1700_tempfit}.
The redshift range $2.58 < z < 2.74$ is chosen since the quality of the FUSE data in the corresponding spectral range is best for both lines of sight.
In the following we will discuss the results for this regime.

\begin{figure}
  \centering
  \resizebox{\hsize}{!}{\includegraphics[bb=35 570 465 740,clip=]{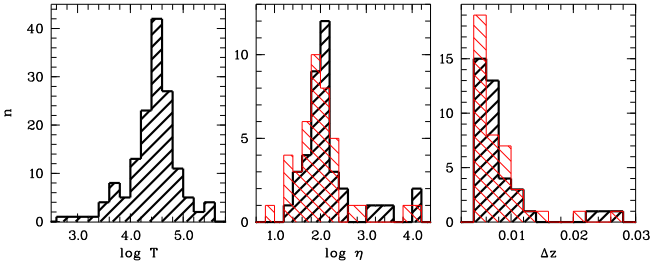}}
  \caption{Distribution of the estimated parameters in the redshift range $2.58 < z < 2.74$ towards HE~2347-4342 and HS~1700+6416.
Presented are the distributions of the line temperature $\log T$ (left panel), the \ion{He}{ii}/\ion{H}{i} column density ratio $\log \eta$ per scale (middle panel, thick line), and the scale size $\Delta z$ of $\eta$ variations (right panel, thick line).
The thin-lined histograms in the right panels indicate the distributions of $\log\eta$ and $\Delta z$, respectively, in the case of turbulent line widths $b_{\ion{He}{ii}} = b_{\ion{H}{i}}$.
  }
  \label{combined_distr}
\end{figure}

As can be seen from the lower panels of Figs.\ \ref{he2347_tempfit} and \ref{hs1700_tempfit}, the estimated temperature is only poorly constrained.
Temperatures can be estimated for about 55\,\% of the lines (presented as circles).
However, in most of the cases (65.5\,\%) the $1\,\sigma$ uncertainties cover the whole range from the high temperature cut-off, when $b_{\ion{H}{i}}$ is interpreted as completely thermal, to the pure turbulent limit.
On the other hand, 45\,\% of the lines favor turbulent line broadening or yield a very low temperature (we set the minimum temperature to $10\,\mathrm{K}$ corresponding to a thermal line width $< 0.5\,\mathrm{km\,s}^{-1}$ which can be interpreted as purely turbulent).
The upper limits estimated for those absorbers are also indicated in Figs.\ \ref{he2347_tempfit} and \ref{hs1700_tempfit}.
In total only 6 lines ($\sim$ 2\,\%) lead to well-constrained temperature estimates.
All of them favor the purely thermal interpretation of the \ion{H}{i} Doppler parameter. 
The resulting distribution of the line temperatures is shown in the left panel of Fig.\ \ref{combined_distr}, excluding components that prefer turbulent line widths.
The average value is $\log T = 4.41 \pm 0.46$.

\begin{figure}
  \centering
  \resizebox{\hsize}{!}{\includegraphics[bb=40 50 485 370,clip=]{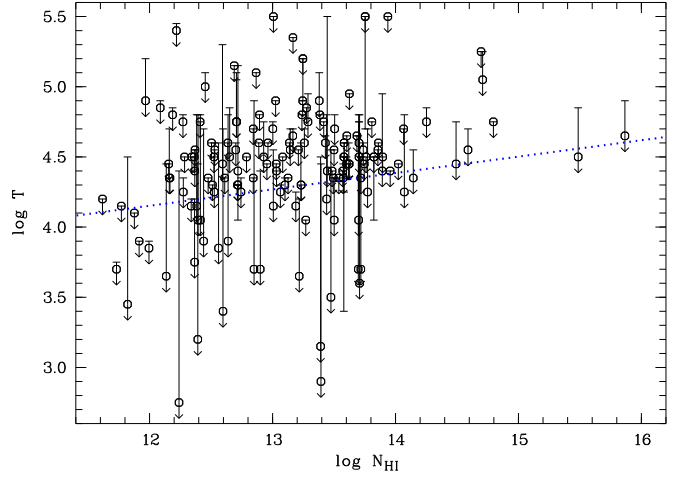}}
  \caption{Line temperature $\log T$ versus \ion{H}{i} column density $\log N_{\ion{H}{i}}$ for the combined sample of HE~2347-4342 and HS~1700+6416 in the redshift range $2.58 < z < 2.74$.
Only lines that lead to realistic best-fit temperatures are indicated.
The mean redshift of the considered line sample is $z = 2.663 \pm 0.082$.
The dotted line indicated a temperature-density relation with $\gamma = 1.17$ and $\log T_0 = 4.23$ as interpolated from the results of \citet{schayeetal2000}.
  }
  \label{logT_logN_combined}
\end{figure}

In order to obtain an estimate of the equation of state the derived temperature is plotted versus the \ion{H}{i} column density in Fig.\ \ref{logT_logN_combined}.
Due to the large confidence intervals of the temperature only a slight correlation between the $\log N_{\ion{H}{i}}$ and $\log T$ can be seen.
The Spearman rank-order correlation coefficient is $r_s = 0.22$.
Consequently, a linear fit to the data according to Eq.\ (\ref{logNHI_logT}) is rather meaningless.
The dotted line in Fig.\ \ref{logT_logN_combined} indicates a temperature-density relation with the parameters $\gamma = 1.17$ and $\log T_0 = 4.23$ as extrapolated from the results of \citet[][see previous section]{schayeetal2000}.
Our temperature estimates have huge error bars (the majority of
points presented in Fig.\ \ref{logT_logN_combined} are in fact upper
limits), but the relation derived from the \citet{schayeetal2000}
results is not inconsistent with our measurements.
Harder constraints of the line temperature are needed to improve the results.
Also the direct temperature estimate would benefit from higher $S/N$ data of the \ion{He}{ii} Ly$\alpha$ forest.

Another point is the determination of the column density ratio $\eta$ and its variation with redshift for the pure turbulent broadening versus the thermal plus turbulent broadening cases.
The right panels of Fig.\ \ref{combined_distr} show the distribution of the resulting $\eta$ values and the scales of its variation.
The average \ion{He}{ii}/\ion{H}{i} ratio for the combined sample is $\log\eta = 2.16 \pm 0.62$ (median $2.01$), where the line of sight towards HE~2347-4342 contributes most of the scatter.
Towards this QSO we find the mean value $\log\eta = 2.21 \pm 0.82$ (median $1.88$; note the outliers in Fig.\ \ref{he2347_tempfit}) while HS~1700+6416 yields $\log\eta = 2.11 \pm 0.32$ (median $2.03$). 
Also shown in Fig.\ \ref{combined_distr} is the distribution of $\eta$ values for the case of fixed $b_{\ion{He}{ii}} = b_{\ion{H}{i}}$ Doppler parameters.
The pure turbulent fits yield roughly $0.1 - 0.2\,\mathrm{dex}$ lower $\eta$ values.
For the combined sample we find the mean $\log\eta = 1.98 \pm 0.59$ (median $1.90$).
The results for the individual sight lines are $\log\eta = 2.00 \pm 0.63$ (median $1.85$) towards HE~2347-4342 and $\log\eta = 1.97 \pm 0.58$ (median $1.95$) towards HS~1700+6416.
Furthermore, the distribution of $\eta$ values is broader in comparison to the thermal fit result.
Fitting a Gaussian to the $\log\eta$ distributions shown in the middle panel of Fig.\ \ref{combined_distr} yields a FWHM of $0.67 \pm 0.12$ for the temperature fit but $0.98 \pm 0.14$ in case of pure turbulent broadening.
An explanation for this finding may be the correlation of $\eta$ with the \ion{H}{i} column density due to the assumption of pure turbulent broadening which we will discuss in the following.
If the estimate of $\eta$ is on average too low for high density absorbers, the distribution of $\eta$ values will become broader and shifted to lower values.

One of the main conclusion of \citet{fechnerreimers2006} has been that the assumption of pure turbulent line broadening may lead to the observed correlation between the strength of the \ion{H}{i} absorption and $\eta$ \citep[e.g.][]{shulletal2004}.
Comparing the results of the fits considering the line temperature or assuming turbulent line widths, we find indications that this correlation is less pronounced if thermal line broadening is taken into account.
The average $\eta$ value for $N_{\ion{H}{i}} > 10^{13}\,\mathrm{cm}^{-2}$ absorbers is $\log\eta = 2.02 \pm 0.55$ (median $2.01$) for the thermal fit and $\log\eta = 1.83 \pm 0.36$ (median $1.87$) in the turbulent case.
If $b_{\ion{He}{ii}} = b_{\ion{H}{i}}$ is assumed, the median $\eta$ value clearly depends on the \ion{H}{i} column density of the absorber, while for $12.0 < \log N_{\ion{H}{i}} < 13.0$ the median $\log\eta$ is $1.93$, it is lower by $0.06\,\mathrm{dex}$ for high column density lines ($\log N_{\ion{H}{i}} > 13.0$).
On the other hand the mean $\eta$ value is independent of \ion{H}{i} column density if thermal broadening is considered. 
In this case the median is $2.01$ in both column density ranges.

The size of the scale of $\eta$ variations is comparable towards both quasars.
The scales are in the range $0.0041 \le \Delta z \le 0.0263$ (mean $\Delta z = 0.0081 \pm 0.0054$, median $0.0061$) corresponding to $\sim 3 - 4\,h^{-1}\,\mathrm{Mpc}$ up to $20\,h^{-1}\,\mathrm{Mpc}$ comoving, which is slightly less than the scales found by \citet{fechnerreimers2006} using the original spectrum fitting method.
As also can be seen from the right panel of Fig.\ \ref{combined_distr}, we find no significant deviation for the distribution of scale sizes derived under the assumption of pure turbulent line broadening or thermal plus turbulent broadening.

\section{Summary and conclusions}\label{conclusions}

We have introduced an approach to take into account the thermal state of the IGM when analyzing the \ion{He}{ii} Ly$\alpha$ forest which has been resolved towards the quasars HE~2347-4342 and HS~1700+6416.
The intention of this procedure is twofold.
At first, systematic errors due to the assumption of pure turbulent line broadening in the analysis of the \ion{He}{ii} Ly$\alpha$ forest should be eliminated.
Following \citet{fechnerreimers2006} the \ion{He}{ii}/\ion{H}{i} column density ratio $\eta$ may be underestimated in case of strong absorbers if turbulent line widths are assumed.
This may lead to a spurious correlation of strength of the absorber with the $\eta$ value.
Secondly, investigating the combined \ion{He}{ii} and \ion{H}{i} Ly$\alpha$ forest could provide in principle an independent strategy to estimate the temperature and the thermal state of the IGM.

The procedure is based on the spectrum fitting method introduced by \citet{fechnerreimers2006}.
Instead of the observed optical data a spectrum computed from the Doppler profile-fitted \ion{H}{i} line list is fitted to the \ion{He}{ii} data.
Using Doppler profiles enables us to modify the temperature of the absorber as an additional parameter.
Consequently, three parameters are estimated, $\gamma$ and $\log T_0$ of the temperature-density relation as well as the column density ratio $\eta$.

Artificial data are used to test the procedure.
We find that a minimum quality of the \ion{He}{ii} data, $S/N \gtrsim 20$, is needed to derive reasonable constraints of IGM's thermal state.
Furthermore, the \ion{H}{i} line list may severely influence the results.
Ambiguities in the decomposition of blends can introduce systematic errors to the resulting fit and even one missing component may prevent the procedure from recovering the true temperature-density relation.

Applying the procedure to the observed spectra of the quasars HE~2347-4342 and HS~1700+6416 leaves the parameters of the temperature-density relation unconstrained due to the low signal-to-noise ratio of the \ion{He}{ii} data ($S/N \sim 5$).
However, some of the fitted redshift ranges show indications of favoring either pure thermal or pure turbulent line broadening.

Taking into account in addition thermal broadening leads to different line widths for \ion{H}{i} and \ion{He}{i} with $b_{\ion{He}{ii}} < b_{\ion{H}{i}}$.
Therefore, the \ion{He}{ii} column density resulting from the fit considering a combination of thermal and turbulent broadening is systematically increased in comparison to a fit assuming the pure turbulent case, i.e.\ $b_{\ion{He}{ii}} = b_{\ion{H}{i}}$.
We find a systematic offset of roughly $0.05\,\mathrm{dex}$ which is within the $1\,\sigma$ uncertainties of the estimates.
The lower the $\chi^2$ of the fit including the temperature-density relation in comparison to the $\chi^2$ of the pure turbulent fit, the higher is the $\eta$ value.
Considering the combined sample of selected fit intervals towards
HS~1700+6416 and HE~2347-4342 the median $\log\eta = 1.96$ is obtained
from the thermal state fit, slightly higher than the median value 1.93 found by \citet{fechnerreimers2006} considering only lines with
$N_{\ion{H}{i}} < 10^{13.0}\,\mathrm{cm}^{-2}$, whereas the turbulent fit leads to $1.88$.

Alternatively, we estimate the thermal line width of the individual absorption features more directly by optimizing the Doppler parameter of the \ion{He}{ii} lines.
In practice the values of the line temperature, the $\eta$ value, and the scale of the variation of $\eta$ are fitted simultaneously.
We find that 45\,\% of the lines favor turbulent broadening.
The inferred temperatures of the remaining 55\,\% of the components suffer from large uncertainties.
Thus, it is impossible to derive a temperature-density relation.
However, comparing the fit which includes a temperature optimization, to a fit assuming pure turbulent line broadening confirms the suspicion of \citet{fechnerreimers2006} that neglecting thermal line widths leads to a correlation of the inferred $\eta$ value with \ion{H}{i} column density.
While the pure turbulent fit results in lower $\eta$ values for high column density absorbers (median $\log\eta = 1.87$ for lines with $\log N_{\ion{H}{i}} > 13.0$ in comparison to $1.93$ for lines with $12.0 < \log N_{\ion{H}{i}} < 13.0$), the thermal fit yields no correlation at all (median $\log\eta = 2.01$ for both column density ranges).
This effect is also noticeable when comparing the distribution of $\eta$ values since a narrower distribution by 46\,\% is found if a temperature is taken into account.

The inferred scales of $\eta$ variations are roughly $4\,h^{-1}\,\mathrm{Mpc}$ comoving.
But also large scales up to $\Delta z \sim 0.03$, corresponding to $\sim 20\,h^{-1}\,\mathrm{Mpc}$ comoving, have been inferred.
The derived scales appear to be independent of the assumption of the dominating broadening mechanism.

\begin{table}
  \caption[]{Summary of $\eta$ values derived with the method of
  direct temperature estimates in the redshift range $2.58 \le z \le 2.74$.
}
  \label{summary}
  $$
  \begin{array}{l c c c c}
    \hline\hline
    \noalign{\smallskip}
              & \multicolumn{2}{c}{\mathrm{turbulent}} & \multicolumn{2}{c}{\mathrm{thermal+turbulent}}\\
              & \langle\log\eta\rangle & \mathrm{median} & \langle\log\eta\rangle & \mathrm{median}\\
            \noalign{\smallskip}
            \hline
            \noalign{\smallskip}
\mathrm{HE~2347-4342} & 2.00 \pm 0.63 & 1.85 & 2.21 \pm 0.82 & 1.88 \\
\mathrm{HS~1700+6416} & 1.97 \pm 0.58 & 1.95 & 2.11 \pm 0.32 & 2.03 \\
\mathrm{both~sight~lines} & 1.98 \pm 0.59 & 1.90 & 2.16 \pm 0.62 & 2.01 \\
            \noalign{\smallskip}
            \hline
         \end{array}
     $$
   \end{table}

Thus, we confirm the existence of a thermal component which results in narrower \ion{He}{ii} features in comparison to \ion{H}{i}, affects the inferred $\eta$ value.
An increased \ion{He}{ii}/\ion{H}{i} ratio is found if a thermal component is taken into account.
At redshifts $2.58 < z < 2.74$ we find $\eta \approx 100$ in agreement
the value found by \citet{fechneretal2006b} at the same redshifts
towards HS~1700+6414 alone 
(a summary of the derived $\eta$ values using the method presented in
  Section \ref{directestimate} is given in Table \ref{summary}).
However, constraining the parameters of the temperature-density relation and thereby providing an independent estimate of the thermal state of the IGM is impossible with the present day quality of \ion{He}{ii} Ly$\alpha$ forest observations.
In order to obtain significant constraints a signal-to-noise ratio of $S/N \gtrsim 20$ would be required which is 4 times higher than the $S/N$ of the data available nowadays.

\begin{acknowledgements}
This work has been supported by the Deutsche Forschungsgemeinschaft (DFG) under RE 353/49-1.
\end{acknowledgements}

\bibliographystyle{aa}
\bibliography{eos}
 
\end{document}